\documentclass[showpacs,showkeys,prl,twocolumn]{revtex4}
\usepackage{graphicx,setspace}
\usepackage{bm}
\begin{document}
\title{ Driven transverse shear waves in a strongly coupled dusty plasma}
\author{P. Bandyopadhyay\footnote[1]{corresponding author.\\ \textit{E-mail address:} banerjee$\_$pintu2002@yahoo.com (P. Bandyopadhyay)\\\textit{Phone No:}+917923962030\\\textit{Fax No:}+917923962277}, G. Prasad, A. Sen and P. K. Kaw}
\address {Institute for Plasma Research, Bhat, Gandhinagar - 382428, India}
\date{\today}
%#####################################################################################
\begin{abstract}
The linear dispersion properties of transverse shear waves in a strongly coupled dusty plasma are experimentally studied in a DC discharge device by exciting them in a controlled manner with a variable frequency external source. The dusty plasma is maintained in the strongly coupled fluid regime with ($1 < \Gamma << \Gamma_c$) where $\Gamma$ is the Coulomb coupling parameter and $\Gamma_c$ is the crystallization limit. A dispersion relation for the transverse waves is experimentally obtained over a frequency range of 0.1 Hz to 2 Hz and found to show good agreement with viscoelastic theoretical results.
\end{abstract}
\maketitle
%#############################################################################
\section{Introduction}
A dusty plasma is a suspension of micron sized particles in a conventional electron-ion plasma. The dust particles get highly charged and hence get strongly coupled to each other through their mutual Coulomb interaction. The strength of interaction is characterized by the Coulomb coupling parameter $\Gamma$ ($=\frac{Q_d^2}{4\pi\epsilon_0d T_d}exp(-\frac{d}\lambda_p))$, where $Q_d$, $d$, $T_d$ and $\lambda_p$ are the dust charge, interparticle distance, dust temperature and plasma Debye length respectively. When $\Gamma$ exceeds a critical value $\Gamma_c$ the dust component can crystallize and behave like an ordered solid. However, even when $\Gamma < \Gamma_c$ but $\Gamma >> 1$, so that the dust component is in a fluid state, the strong correlations can lead to the development of short range order among the dust particles and endow it with some solid-like properties. A consequence of such viscoelastic behaviour is the possibility of exciting and sustaining transverse shear waves in the medium - a novel collective phenomenon that cannot take place in a normal fluid. Wave propagation in strongly coupled media have been investigated in a number of past theoretical studies \cite{rosen,wang1,sen,sen1,kalman1,wang2,kalman2}. Rosenberg and Kalman \cite{rosen} looked at its collective modes using the Quasi-localized charge approximation (QLCA). Wang and Bhattacharjee \cite{wang1} used the \lq\lq generalized thermodynamic'' approach of Kadanoff and Martin \cite{kada}.
Kaw \& Sen \cite{sen} adopted a more phenomenological approach and employed the Generalized Hydrodynamic (GH) Model to derive the linear dispersion relation for longitudinal dust acoustic modes as well as for transverse shear modes. The existence of transverse modes has also been predicted from numerical simulation studies carried out by Ohta and Hamaguchi \cite{ohta} and subsequently also derived within the framework of the QLCA by Kalman and co-workers \cite{kalman1,kalman2}. Although transverse shear waves are relatively easy to excite and study in ordered crystalline media \cite{nuno1,misawa,nuno2} their experimental detection in the fluid regime is more challenging. For dusty plasmas the first experimental observation of transverse shear waves in the fluid regime was made by Pramanik {\it et al} \cite{pramanik}  who looked at self excitations of such waves in conjunction with longitudinal dust acoustic waves in a laboratory device. By varying the neutral gas pressure they were able to vary in a limited fashion the natural frequencies of the waves and thereby experimentally determine the dispersion relation. 
However in their experiment there was no direct control over the frequency of the excitations nor a direct measurement of it. Further, by changing the neutral pressure one also changes some of the equilibrium parameters which leads to an additional complexity in the accurate determination of the dispersion relation. To obviate these difficulties and to further critically test and confirm the experimental findings of \cite{pramanik} we have carried out a set of controlled experiments in which the shear waves are excited by an external source whose frequency can be varied over a range. Although many such controlled driven experiments have been carried out in the past for longitudinal dust acoustic waves \cite{barkan,thompson,pieper,pintu}, there have been very few attempts to study controlled excitations of transverse shear waves in a strongly coupled plasma. A notable exception is the work of Piel, Nosenko and Goree \cite{piel} who excited shear waves in a single layer of suspended particles in the solid and liquid phases by external perturbations with a laser. In our experiment we have looked at shear waves in a three-dimensional dusty plasma cloud configuration under conditions similar to that reported in \cite{pramanik}. Our dispersion curve obtained by changing the applied frequency over a finite range is found to agree quite well
with the GH model predictions and thus provide an added confirmation of the previous results of Pramanik {\it et al} \cite{pramanik}.
\section{Experimental Set-up}
The experiment was carried out in the set-up shown in Fig.~1. The SS cylindrical chamber has eight radial ports and two axial ports. The radial ports are used for pumping, introducing gas, feeding power to the electrodes, diagnostics, etc. The axial ports are used only for optical diagnostics for the dust particles. All the unused ports were covered by nylon bushes to avoid any hollow cathode glow discharge. The vacuum chamber was pumped down to a base
%===============================================================================
\begin{figure} 
\includegraphics[width=0.450\textwidth]{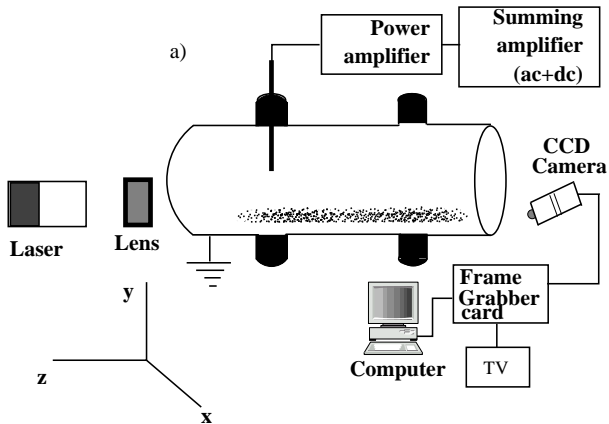}
\caption{The schematic of experimental setup}
\label{fig:chamber}
\end{figure}
%=======================================================================================
pressure of $10^{-3}$ mbar by a rotary pump. It was then purged with argon gas and subsequently pumped down again to the base pressure. This process was repeated several times by using a precision needle valve. Then micron sized kaolin dust particles were sprinkled at the bottom of the vacuum chamber. The pressure was kept constant at 1 mbar. The entire chamber was covered by a thin SS sheet to avoid stray arcing. A discharge was produced between a rod shaped anode and the grounded vessel (used as a cathode) at an applied voltage of $V_a=600$ volt. The applied voltage was then reduced to $542$ volts and the neutral gas pressure was gradually reduced to $0.09$ mbar to achieve a dense dust cloud. \par
The levitated dust particles were illuminated by a green Nd-Yag diode laser ($\lambda_{laser} = 532 \mu m $) light. The laser light was spread into a sheet by a cylindrical lens and the forward scattered light from the dust cloud was used to visualize the dust particles. In these set of experiments the cylindrical lens was kept vertical to illuminate the $y-z$ plane of the dust column. The scattered light from the dust particles was captured using a CCD camera (25 fps) which was kept at an angle of $15^0$ to this plane. Further, the wavelength measurements in the vertical plane ($y-z$) were duly corrected to account for the geometrical effects arising from the small but finite angle of inclination of the CCD camera with respect to the $z$ axis. Video frames were digitised by a frame grabber card with eight bits of intensity resolution and recorded in a high speed computer.\par
An AC signal ($V_{amp}= 85$ volts using a signal generator and a power amplifier) was superimposed on to the discharge voltage to excite oscillations in the dust cloud (see Fig.~2). The mean discharge current was $105$ mA at $542$ volts. The frequency of the applied AC signal was varied from 0.1 to 2 Hz during the experiment to excite acoustic and transverse waves.
\par
The plasma parameters like the ion density ($n_i$) and the electron temperature ($T_e$) were measured initially by using a single Langmuir probe in the absence of dust particles inside a plasma. The electron density was estimated from the modified quasi-neutrality conditions of the dusty plasma. A hot emissive probe was used to measure the plasma potential and the floating potential. The radial profiles of the floating and the plasma potentials are shown in Fig.~3. The dust temperature ($T_d$) was calculated from the velocity of the particles by tracing single particle trajectories in different frames. The wave length of the driven dust acoustic wave was obtained by measurements made on the still images.
%===================================================================================== 
\begin{figure} 
\includegraphics{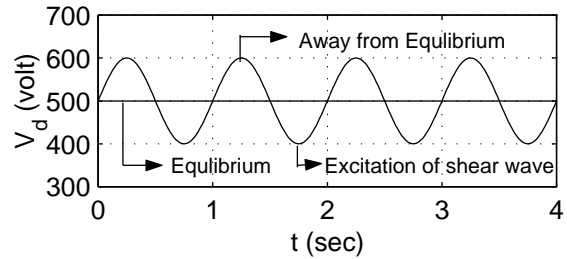}
\caption{Typical AC modulation voltage superimposed over the DC discharge voltage to excite transverse shear waves.}
\label{fig:mod_vol}
\end{figure}
%=======================================================================================
%===================================================================================== 
\begin{figure} 
\includegraphics[width=0.4\textwidth]{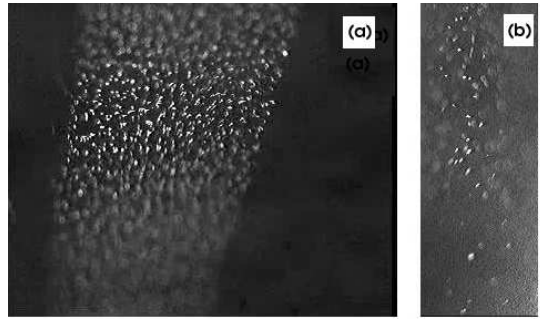}
\caption{Dust particle levitation in (a) $x-z$ and b) $y-z$ plane.}
\label{fig:equi}
\end{figure}
%=======================================================================================
\section{Results and Discussion}
In the initial stage of the experiment, an equilibrium dust cloud consisting of layers of micron sized kaolin particles was observed to form in the sheath region of the discharge. The layer heights are governed by the force balance conditions acting on the particles and are determined by the discharge parameters. Fig.~4(a) shows a horizontal view ($x-z$) of the equilibrium dust cloud and Fig.~4(b) shows the same in the ($y-z$) plane. For the vertical levitation of the dust cloud, the upward electrostatic force $F_E$ based on the inhomogeneous sheath electric field suspends negatively charged dust particles against the downward gravitational force $F_g$. For the micron sized dust particles, the other forces (like neutral drag force and thermophoretic force) do not contribute significantly as they are smaller than these two major forces ($F_E$ and $F_g$) by an order of magnitude. The radial electric field provides the radial confinement of the particles. It is clear from Fig.~4(b) that the smaller particles levitate to the top surface of the cloud while the big particles reside at the bottom indicating the existence of a mass and a charge gradient in the $y$ direction. Such a gradient is a consequence of the potential gradient existing in the sheath region and the dispersion in the particle sizes.\par
%===================================================================================== 
\begin{figure} 
\includegraphics{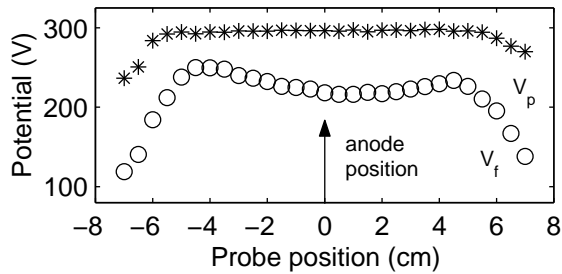}
\caption{Variation of Plasma Potential ($V_p$) and Floating potential ($V_f$) with radial probe position.}
\label{fig:potentials}
\end{figure}
%=====================================================================================
Soon after the formation of the equilibrium dust cloud, at $P=0.09$ mbar, $V_a=542$ volts and $I_d=105$ mA, spontaneous excitation of longitudinal oscillations were observed to occur, and are shown in Fig.~5(a). These oscillations, which are basically dust acoustic oscillations, were seen to propagate in the $-y$ direction (in the direction of the sheath electric field) with a phase velocity $V_{ph}\sim 4.5$ cm/sec (which was measured by tracking a single crest in consecutive frames). In the present set of experiments, the typical plasma parameters are $n_i\sim3\times10^{15}$ m$^{-3}$, $T_e=8-10$ eV, and $T_i=0.03$ eV. The dusty plasma parameters are $r \sim 2$ $\mu$m, $d \sim 100$ $\mu$m, $T_d\sim 5-7$ eV, $n_d\sim2.4\times10^{11}$ m$^{-3}$, $Z_d\sim2.8\times10^4$ and $M_d\sim1.1\times10^{-13}$ kg. The high dust particle temperature noticed in the present set of experiments was also reported by Melzer \textit{et al.} \cite{melzer}. Using the above parameters we get an average value of $\lambda_p \sim 23.5$ $\mu$m and $\omega_{pd}\sim 2.3\times10^3$ Hz. In the limit $k\lambda_p<<1$, the theoretically calculated phase velocity for dust acoustic waves, $C_{da} \sim \lambda_{p}\omega_{pd}\sim5.3$ cm/sec which is in close agreement with the experimental measurements.\par
%===================================================================================== 
\begin{figure} 
\includegraphics{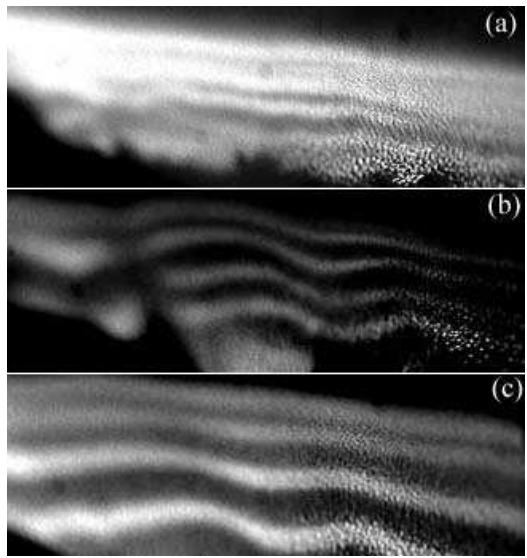}
\caption{a) Background oscillation of dust acoustic waves in the absence of a modulation voltage. Typical transverse shear waves for the applied frequency b) 0.690 Hz and c) 0.486 Hz}
\label{fig:waves}
\end{figure}
%=====================================================================================
Now when a modulation voltage of a very low frequency ($\sim$ few Hz) was superimposed over the DC discharge voltage an additional wave motion was seen to appear in the dust cloud over the spontaneously excited dust acoustic wave (see Fig.~5(b) and Fig.~5(c)). It was also noticed that the phase velocity of the acoustic oscillation remained unaffected with the application of the modulation voltage. The additional wave motion was observed to propagate perpendicularly to the direction of propagation of the acoustic waves. The direction of propagation of these new additional waves was also seen to be perpendicular to the direction of the particle oscillation and hence these waves are truly transverse waves. Fig. 5(b) and Fig.~5(c) show the typical acoustic and shear waves of different wave lengths ($f = 0.690$ Hz and $f = 0.486$ Hz). The frequency of the applied modulating voltage was varied over a finite range to get the dispersion relation which is shown in Fig.~6. The average phase velocity was measured as $\sim 7.9$ mm/sec. In the present set of experiments the excitation of pure transverse  waves was not possible and we were able to excite them only over the dust acoustic oscillations, as was observed earlier in the self excitation case \cite{pramanik}. This is an interesting issue that is not fully understood yet and needs to be explored in greater depth and addressed in future studies.\par
To compare our experimental results to past theoretical predictions we have used the viscoelastic theory proposed by Kaw and Sen  \cite{sen}. According to this theory in the limit $\omega\tau_m>>1$ ($\tau_m$ is the relaxation time), the dispersion relation for the transverse shear mode is given by:
%$$$$$$$$$$$$$$$$$$$$$$$$$$$$$$$$$$$$$$$$$$$$$$$$$$$$$$$$$$$$$$$$$$$$$$$$$$$$$$$$
\begin{equation}
\omega^2=\frac{3}{4}\frac{k^2T_d}{M_d}\left(1-\gamma_d\mu_d+\frac{4}{15}u\right),
\label{eqn:eqn1}
\end{equation} 
%$$$$$$$$$$$$$$$$$$$$$$$$$$$$$$$$$$$$$$$$$$$$$$$$$$$$$$$$$$$$$$$$$$$$$$$$$$$$$$$$
where $\gamma_d$ and $\mu_d$ are the adiabatic index and compressibility of the medium and $u$ is the excess potential energy. The values of $u$ and $\mu_d$ can be estimated from $\Gamma$ according to the following relations,
%$$$$$$$$$$$$$$$$$$$$$$$$$$$$$$$$$$$$$$$$$$$$$$$$$$$$$$$$$$$$$$$$$$$$$$$$$$$$$$$$
\begin{eqnarray}
\mu_d&=&\frac{1}{Td}\left(\frac{\partial P}{\partial n}\right)_T=1+\frac{u(\Gamma)}{3}+\frac{\Gamma}{9}\frac{\partial u(\Gamma)}{\partial \Gamma}
\nonumber\\
u(\Gamma)&=&-0.89\Gamma+0.95\Gamma^{1/4}+0.19\Gamma^{-1/4}-0.81.
\nonumber
\end{eqnarray}
%$$$$$$$$$$$$$$$$$$$$$$$$$$$$$$$$$$$$$$$$$$$$$$$$$$$$$$$$$$$$$$$$$$$$$$$$$$$$$$$$
\indent Eqn.~\ref{eqn:eqn1} shows the theoretical linear dispersion relation which is plotted by a solid line in Fig.~6. The value of $\Gamma$ can be estimated from the given experimental parameters. As mentioned earlier, the measured inter-grain distance $ d \sim 100$ $\mu$m which makes the screening factor $exp(-d/\lambda_p)$ to have a value of $\sim 0.0142 $. The unscreened contribution due to the factor $Q_d^2/4\pi\epsilon_0dT_d$ is about $ \sim 1600$, making the Coulomb coupling parameter to have a value $\sim$ 23. From the calculated value of $u$ and $\mu_d$, the phase velocity of the shear wave according to the GH model (viscoelastic theory) (Eqn.~\ref{eqn:eqn1}) comes out to be $V_{ph} \approx 7.6$ mm/sec. This is in good agreement with the above experimental results.\par
It is appropriate to mention here that there also exists an excitation threshold for sustaining transverse shear waves in a dusty plasma. Basically, as discussed in \cite{nuno1}, dust particles constituting the shear wave gain energy from the inhomogeneous sheath electric field while oscillating in the upward direction and lose their energy due to the collisions with the neutral gas molecules during the time of return. Excitation of transverse shear waves therefore requires that the energy gain exceeds the energy loss. This condition results in a threshold limit on the plasma density and the neutral gas
%===================================================================================== 
\begin{figure} 
\includegraphics{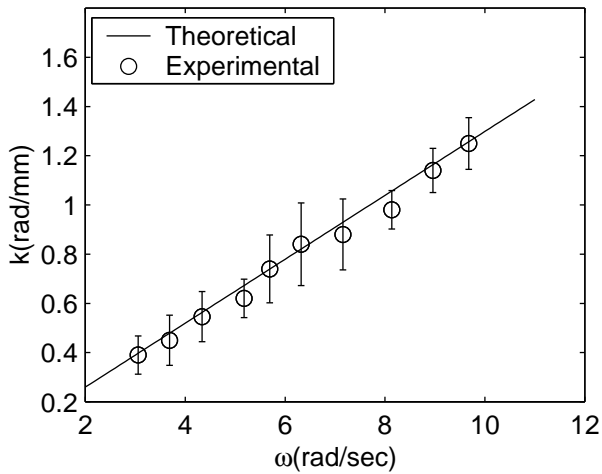}
\caption{ The experimental (empty circle) and theoretical (solid line) dispersion relation of transverse shear wave.}
\label{fig:disp_reln}
\end{figure}
%=======================================================================================
pressure. Kalman \textit{et al}. \cite{kalman1} in their theoretical calculations have also predicted that the transverse mode  cannot be excited in a regime where $\nu_{dn}/\omega_{pd} > 0.09$. They have also suggested that the excitation of transverse mode can be facilitated by using a lighter gas and a dust material with a larger specific gravity to reduce the collisional frequency at low pressure. In our case the ratio of $\nu_{dn}$ to $\omega_{pd}$ is very low $\sim$ 0.002 and so we have had no difficulty in exciting and sustaining the transverse waves. However for verifying the threshold mechanism we have increased the neutral pressure to P=0.3 mbar where the ratio of $\nu_{dn}$ to $\omega_{pd}$ is higher ($\sim$ 0.09). In such a case we have observed that the transverse oscillations get severely damped and are very difficult to sustain. These measurements lend further support to the phenomenological physical model for the excitation and sustenance of transverse shear modes in a strongly coupled dusty plasma fluid. 
\section{Conclusion}
In this letter, we have reported on experimental observations of externally excited low frequency transverse shear waves in a strongly coupled dusty plasma fluid which in a sense confirms and complements the earlier observations on self-excited waves reported in \cite{pramanik}. The waves were excited by the application of a modulated AC voltage superposed on a DC discharge voltage. The transverse shear waves were found to be excited over a background of low frequency dust acoustic waves. The experimental dispersion relation compares quite favorably with existing visco-elastic model calculations. We also observe a strong damping of the modes at higher neutral gas pressures that is in accord with theoretical predictions regarding the role of dust-neutral collisions in draining energy away from these modes.

\end{document}